\documentclass[lettersize,journal]{IEEEtran}
		\usepackage[bookmarks=false]{hyperref}
\usepackage{amsmath,amsfonts}
\usepackage{algorithm}
\usepackage{algpseudocode}
\usepackage{array}
\usepackage[caption=false,font=normalsize,labelfont=sf,textfont=sf]{subfig}
\usepackage{textcomp}
\usepackage{stfloats}
\usepackage{url}
\usepackage{verbatim}
\usepackage{graphicx}
\usepackage{booktabs}
\usepackage{multirow}
\usepackage{makecell}
\usepackage{adjustbox}
\usepackage{lmodern}
\usepackage[T1]{fontenc}
\usepackage[utf8]{inputenc}
\usepackage{threeparttable} 
\def\BibTeX{{\rm B\kern-.05em{\sc i\kern-.025em b}\kern-.08em
    T\kern-.1667em\lower.7ex\hbox{E}\kern-.125emX}}
\usepackage{balance}
\begin{document}
\title{GDNTT: an Area-Efficient Parallel NTT Accelerator Using Glitch-Driven Near-Memory Computing and Reconfigurable 10T SRAM}
\author{Hengyu Ding$^\dagger$, Houran Ji$^\dagger$, Jia Li, Jinhang Chen, Chiu-Wing Sham,~\IEEEmembership{Senior Member,~IEEE}, Yao Wang,~\IEEEmembership{Member,~IEEE}
\thanks{This work was supported by the Science and Technology Project of Henan Province under Grant 232102211076. \textit{(Corresponding author: Yao Wang)}

$^\dagger$These authors contributed equally to this work.

Hengyu Ding, Houran Ji, Jia Li, Jinhang Chen and Yao Wang are with the School of Electrical and Information Engineering, Zhengzhou University, Zhengzhou 450001, China (e-mail: ieyaowang@zzu.edu.cn).

Chiu-Wing Sham is with the School of Computer Science, University of Auckland, Auckland 1010, New Zealand.
}}

\markboth{Journal of \LaTeX\ Class Files,~Vol.~18, No.~9, September~2020}%
{How to Use the IEEEtran \LaTeX \ Templates}

\maketitle

\begin{abstract}
With the rapid advancement of quantum computing technology, post-quantum cryptography (PQC) has emerged as a pivotal direction for next-generation encryption standards. Among these, lattice-based cryptographic schemes rely heavily on the fast Number Theoretic Transform (NTT) over polynomial rings, whose performance directly determines encryption/decryption throughput and energy efficiency. However, existing software-based NTT implementations struggle to meet the real-time performance and low-power requirements of IoT and edge devices. To address this challenge, this paper proposes an area-efficient highly parallel NTT accelerator with glitch-driven near-memory computing (GDNTT). The design integrates a 10T SRAM for data storage, enabling flexible row/column data access and streamlining circuit mapping strategies. Furthermore, a glitch generator is incorporated into the near-memory computing unit, significantly reducing the latency of butterfly operations. Evaluation results show that the proposed NTT accelerator achieves a 1.5$\sim$28× improvement in throughput-per-area compared to the state-of-the-art.
\end{abstract}

\section{Introduction}
\IEEEPARstart{T}{he} rise of quantum computing threatens traditional encryption methods. As a post-quantum candidate, lattice-based cryptography resists quantum attacks but faces efficiency bottlenecks in polynomial multiplication \cite{Aysu2013, He2023}. To address this, the Number Theoretic Transform (NTT) and its inverse (INTT) are widely adopted to accelerate computations, though they remain the most time-consuming operations in hardware implementations. Thus, optimizing NTT/INTT accelerators has become a critical research focus.

NTT accelerator architectures can be broadly classified into two types: Von Neumann-based and in-memory/near-memory computing. The Von Neumann approach, with its separate compute and memory units, faces significant data movement overhead, limiting latency and memory efficiency \cite{Zhang2021, Paludo2022, Cheng2024}. In-memory/near-memory computing integrates processing units within or near memory, reducing data transfer overhead and improving energy efficiency and performance. Nejatollahi et al. \cite{Nejatollahi2020} proposed a processing-in-memory NTT architecture with fast in-memory multiplication and modulo operations. Li et al. \cite{Li2022} developed a near-memory data mapping technique to optimize butterfly operation storage and resolve data conflicts. Park et al. \cite{Park2022} implemented an RRAM-based CiM NTT accelerator using vector-matrix multiplication for maximal parallelism. Zhang et al. \cite{Zhang2023} proposed a fast in-SRAM NTT with optimized bit-parallel modular multiplication and shift operations. Zhang et al. \cite{Zhang2025} developed an NTT using hybrid redundant numbers for carry-free modular multiplication and optimized memory patterns. Pakala et al. \cite{Pakala2025} introduced the MBSNTT architecture, which combines multi-bit serial modular multiplication with in-memory computing for fully parallel NTT operations. We observe that most in-memory or near-memory computing NTT accelerators require significant memory overhead in exchange for highly parallelized processing. For example, the 1024-point NTT proposed in \cite{Zhao2024}, based on the Von Neumann architecture, uses only 1024 bits of RAM. In contrast, the 1024-point in-memory computing NTT designed in \cite{Zhang2025} utilizes 61440 bits of SRAM, which is 60 times that of the former.

In this brief, we propose GD-NTT (Glitch-Driven NTT), a highly parallel accelerator architecture featuring low memory and area overhead. The 10T SRAM supports concurrent row-column access for efficient butterfly operand pairing, while the glitch-based clock division technique generates precise timing signals to reduce near-memory computation latency. This dual optimization minimizes memory usage and maximizes throughput. Experimental results show the NTT achieves 67.1 kNTT/s at 256-point with only 0.006 $mm^2$ area, providing 1.5$\sim$28× better throughput-per-area compared to the state-of-the-art.

\section{Background}
NTT is a variant of the Fast Fourier Transform (FFT) performed over finite fields. Compared to FFT, NTT operates in a finite field, avoiding the errors and complexity associated with floating-point operations. Its primary function is to accelerate polynomial multiplication. The transformation converts the polynomial coefficients from the time domain to the frequency domain, reducing the computational complexity of polynomial multiplication from $O(N^2)$ to $O(N \log N)$. NTT plays an important role in cryptography, particularly in PQC and fully homomorphic encryption (FHE). NTT's basic principle is to use primitive roots and unity roots on finite fields to achieve fast transforms. Given a finite field \( F_q \) (where \( q \) is a prime), and a primitive root \( \omega \) such that \( \omega^n \equiv 1 \pmod{q} \) and for all \( i < n \), \( \omega^i \pmod{q} \) is an \( n \)-th primitive root of unity, for a polynomial $a(x) = \sum_{j=0}^{n-1} a_j x^j$, its \( n \)-point NTT computation is given by
\begin{equation}
\label{NTT}
A_i = \sum_{j=0}^{n-1} a_j \omega^{ij} \pmod{q} \quad (i = 0, 1, \dots, n-1)
\end{equation}
INTT recovers the original polynomial \( a(x) \) from \( A(x) \), which can be expressed as
\begin{equation}
\label{INTT}
a_i = \frac{1}{n} \sum_{j=0}^{n-1} A_j \cdot \omega^{-ij}
\end{equation}
In this brief, we focus on implementing NTT and INTT transformations using the Cooley-Tukey and Gentleman-Sande algorithms, respectively. The Cooley-Tukey method, as shown in Alg. \ref{CT}, decomposes large-scale NTT operations into smaller sub-operations through multi-layer loops, processing each part iteratively. In each loop, the results of adjacent computations are combined through butterfly operations, enabling efficient parallel computation of all terms. When performing the INTT transformation, the Gentleman-Sande method uses similar butterfly operations to revert the point-value form back to polynomial coefficients, as shown in Alg. \ref{GS}.

\begin{algorithm}
\caption{NTT algorithm with Cooley-Tukey method}
\label{CT}
\textbf{Input:} Polynomial \( x \), modulus \( q \), length \( n \), inverse twiddle factor \( \omega\) \\
\textbf{Output:} \(X=NTT(x)\)
\begin{algorithmic}[1]
\State \(\hat{\omega} \gets \text{bit-reverse}(\omega)\)
\If{n == 1}
    \State \text{return} \(a\)
\EndIf
\For{\(i = 1; i < \log_2 n; i = i + 1\)}
    \State \(m \gets 2^i\)
    \For{\(j = 0; j < (m/2); j = j + 1\)}
        \For{\(k = 0; k < (n/m); k = k + 1\)}
            \State \(t \gets x[k + 2*j*(n/m)]\)
            \State \(u \gets \hat{\omega}^{j}*{x[k + 2*(j+1)*(n/m)]} \mod q\)
            \State \(X[k + 2*j*(n/m)] \gets (t + u) \mod q\)
            \State \(X[k + (2j+1)*(n/m)] \gets (t - u) \mod q\)
        \EndFor
    \EndFor
\EndFor
\end{algorithmic}
\end{algorithm}

\begin{algorithm}
\caption{INTT algorithm with Gentleman-Sande method}
\label{GS}
\textbf{Input:} Polynomial \( X \), modulus \( q \), length \( n \), inverse twiddle factor \( \omega^{-1} \) \\
\textbf{Output:} \(x=INTT(X)\)
\begin{algorithmic}[1]
\State \(\hat{\omega^{-1}} \gets \text{bit-reverse}(\omega^{-1})\)
\State \(\hat{X} \gets \text{bit-reverse}(X)\)
\For{\(i = 1; i < \log_2 n; i = i + 1\)}
    \State \(m \gets 2^i\)
    \For{\(j = 0; j < (n/m); j = j + 1\)}
        \For{\(k = 0; k < m/2; k = k + 1\)}
            \State \(t \gets \hat{X}[k + i*m]\)
            \State \(u \gets \hat{\omega^{-j}} * \hat{X}[k + j*m + m/2] \mod q\)
            \State \(X[k + j*m] \gets (t + u) \mod q\)
            \State \(X[k + j*m + m/2] \gets (t - u) \mod q\)
        \EndFor
    \EndFor
\EndFor
\end{algorithmic}
\end{algorithm}

\section{Proposed Architecture}
\subsection{Overall Architecture}
\begin{figure*}[t]
    \centering
    \includegraphics[width=1\textwidth]{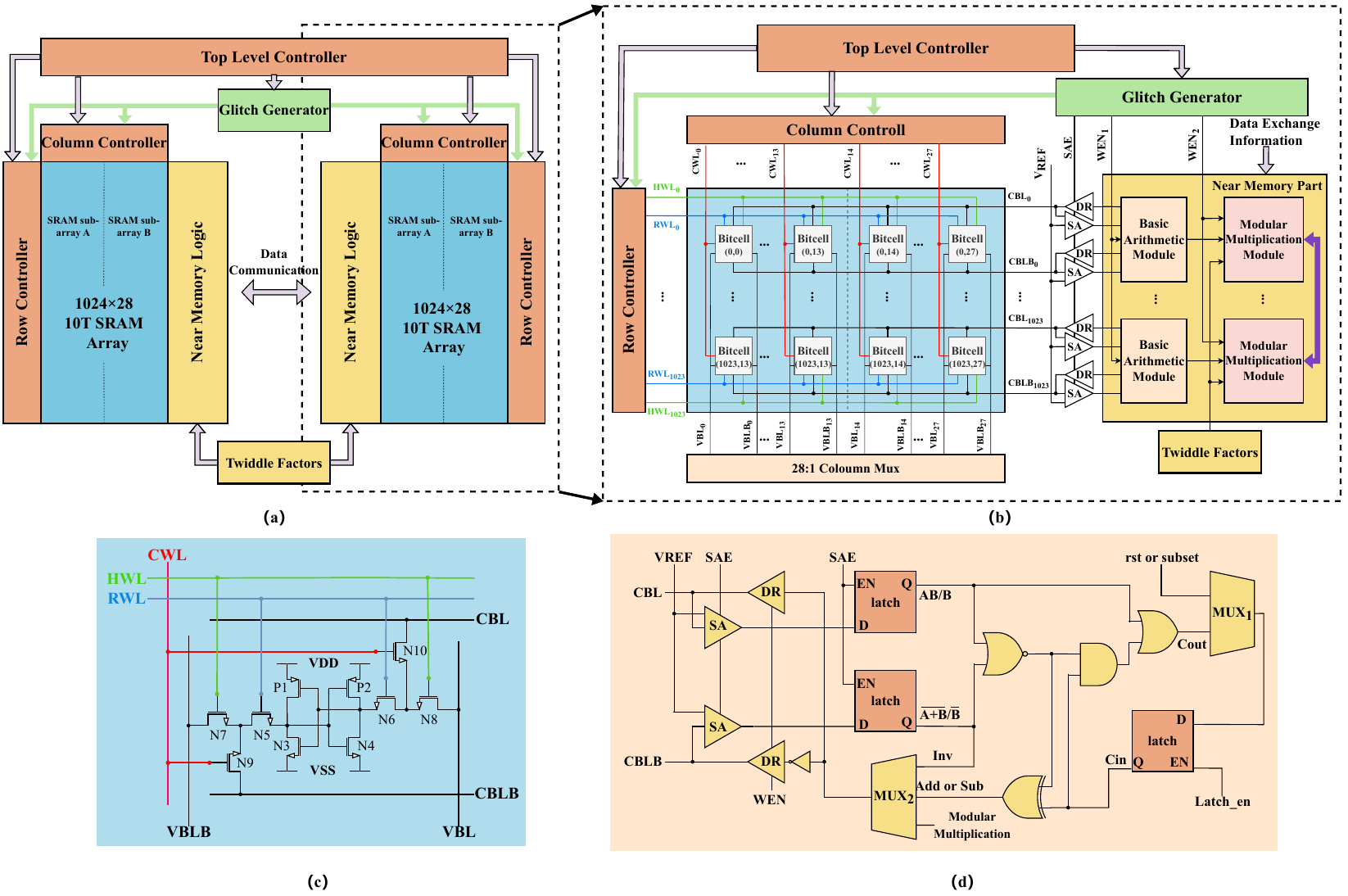}
    \caption{(a) Overall architecture of the proposed NTT accelerator. (b) Detailed structure of the near-memory computation unit. (c) Structure of the proposed 10T SRAM bit-cell. (d) Structure of the basic arithmetic module.}
    \label{fig:overall}
\end{figure*}
The overall architecture of the proposed NTT/INTT near-memory computing accelerator is illustrated in Fig.~\ref{fig:overall}(a). The accelerator consists of modules such as the top level controller, glitch generator, SRAM array, near memory logic, and twiddle factor memory. The top level controller acts as the central control unit, overseeing the operation of all other components, including the SRAM array's row and column control, data retrieval, and writing back of results. The SRAM arrays, represented in blue, are used to store operands and intermediate results in two separate sets. The near memory logic units (highlighted in yellow) carry out computations close to the memory. These modules contain a butterfly operation unit that performs arithmetic operations, such as multiplication, and a twiddle factor memory that stores pre-computed rotation factors used in the butterfly operations. Furthermore, these near memory logic units enable data exchange between the two SRAM arrays, facilitating the point-wise multiplication of two data sets post-NTT computation. The glitch generator generates multiple narrow pulse-width signals with the same frequency as the main clock, used to synchronously drive row control, column control, and near-memory computing units, enabling the continuous execution of multi-step near-memory logic operations within a single clock cycle. Driven by a series of narrow pulse width signals generated by the glitch generator, we are able to complete SRAM data reading, near-memory computation, and storage of the computation results within a single clock cycle.

\subsection{Design of 10T SRAM and Parallel Memory Access Scheme}
In conventional vector computing, there is no logical operation between rows, that is, the rows are decoupled. However, in the NTT calculation process, there is a coupling relationship between rows. Conventional 8T SRAM architectures rely on column word lines (CWL) to control the access of each bit cell in a column. However, precise row-level control is unachievable with this configuration. To address this limitation, we designed a 10T SRAM bit cell, shown in Fig.~\ref{fig:overall}(c). 

In this design, when performing read/write operations via VBL and VBLB, both RWL and HWL are set to high while CWL remains low. This configuration turns on transistors N5, N6, N7, and N8, connecting VBL and VBLB to the bit cell while isolating CBL and CBLB through N9 and N10. In this configuration, the SRAM bit cell can perform read/write operations on the near-memory computing unit via the VBL and VBLB lines. Conversely, when accessing the CBL and CBLB lines, RWL and CWL are pulled to high to achieve precise bit cell control, while HWL is set to low. This activates N5, N6, N8, and N9, linking CBL and CBLB to the circuit while isolating VBL and VBLB via N7 and N8. In this case, the SRAM bit cell is connected to the input/output interface of the NTT accelerator through a column MUX. This approach allows dynamic row selection during data migration, thereby achieving precise matching of butterfly operation data pairs across different computational stages.

\subsection{Near-Memory Computation}
\begin{figure}[t]
    \centering
    \includegraphics[width=0.48\textwidth]{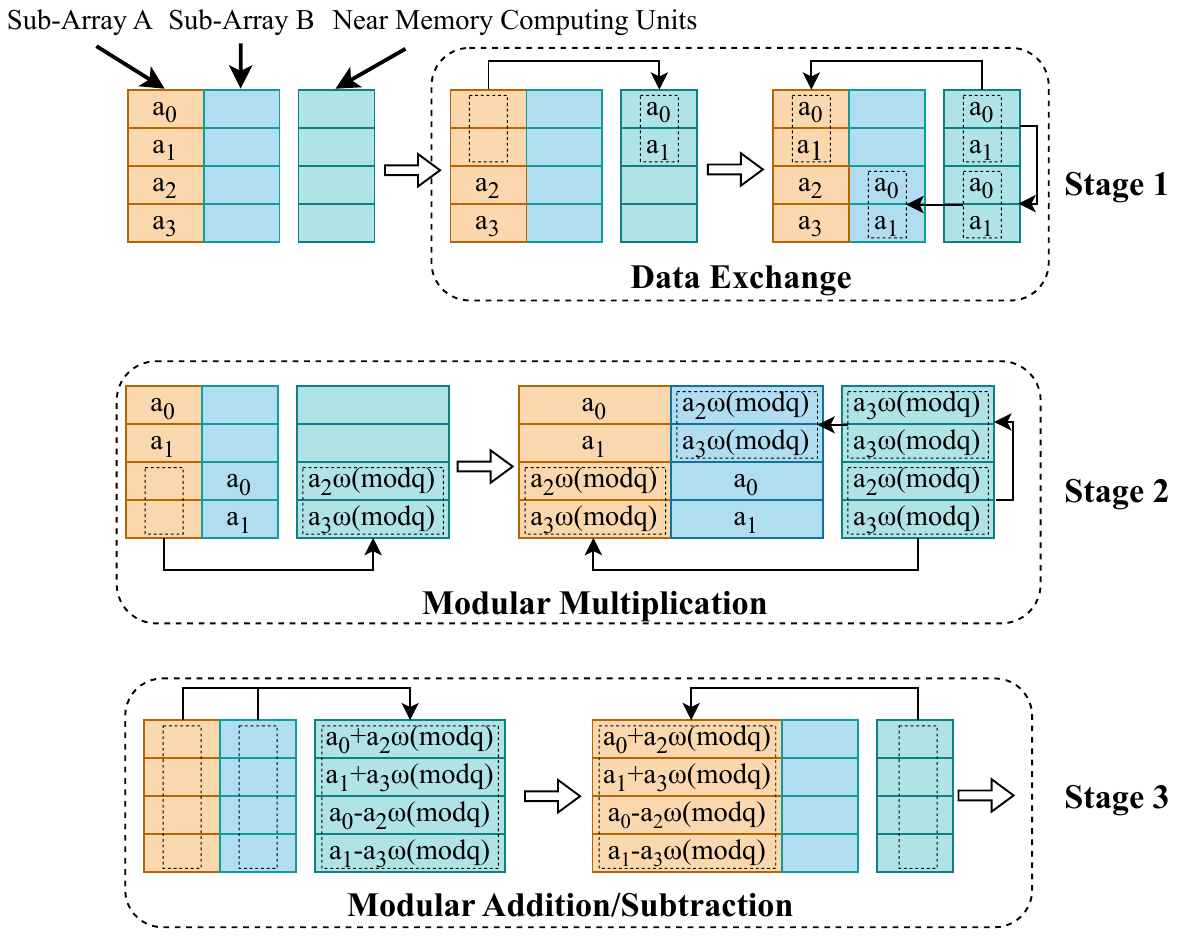} 
    \caption{Detailed NTT caculation flow.}
    \label{fig:caculation}
\end{figure}

 Our NTT/INTT operations are based on the Cooley-Tukey algorithm and the Gentleman-Sande algorithm, implementing the conventional data flow of NTT/INTT computations. Fig.~\ref{fig:caculation} provides an illustrative example of the computational process for a single-stage butterfly operation in a 4-point NTT. The mapping of data between butterfly operation stages is achieved through two data exchanges in $Stage 1$ and $Stage 2$. In $Stage 1$, we perform data migration of the original data to enable direct parallel computation. The process begins by reading corresponding data according to the number of butterfly stages. Once data is read from the SRAM via CBL and CBLB lines, it is amplified by sense amplifiers and fed into the near-memory computation unit. Specifically, $a_0$ and $a_1$ are sequentially loaded into registers in the near-memory computation unit before data transfer. The purple lines in  Fig.~\ref{fig:overall}(b) represent the data transmission channels between different near-memory computing units. Through these data channels, $a_0$ and $a_1$ are transmitted to another set of near-memory computing units, and then the data from the two sets of near-memory computing units are written in parallel into SRAM sub-arrays A and B, respectively.  Since the original data in the SRAM sub-array A will disappear after read operations, a copy of the original data must be rewritten back to the SRAM sub-array A. At this stage, we have successfully completed half of the data exchange process.

The modular multiplication of NTT is demonstrated in $Stage 2$ of Fig.~\ref{fig:caculation}. We employ the standard Barrett modular multiplication method. $a_2$ and $a_3$ are read in parallel from the SRAM array and carries out modular multiplication operations in near-memory computing units. After completing the modular multiplication,  the results are written back to the remaining blank SRAM arrays in sub-array A and B using the identical data exchange methodology implemented in Stage 1.

$Stage 3$ demonstrates the modular addition/subtraction operations. The detailed data transfer mechanism will be systematically elaborated in the subsequent subsection. Fig.~\ref{fig:overall}(d) illustrates the internal structure of the basic arithmetic module, which carries out addition and subtraction operations. When executing addition, operands $A$ and $B$ are fed into latches via CBL and CBLB, where XOR logic computes $A \oplus B$. Carry information is propagated through MUX1, which selects either an initial reset value or the previous stage's carry-out value. The sum is computed iteratively and stored back in SRAM. For subtraction, $B$ is first inverted, then A and the inverted B carry out an addition operation with an initial carry-in value of 1. The MUX2 determines whether the output is from addition/subtraction or the modular multiplication module. Upon completion of Stage 3, the processed data is concurrently written into SRAM sub-array A through parallel write operations, where it is retained for subsequent computational phases.

\subsection{Glitch-driven High-speed Butterfly Operation}

\begin{figure}[t]
    \centering
    \includegraphics[width=0.5\textwidth]{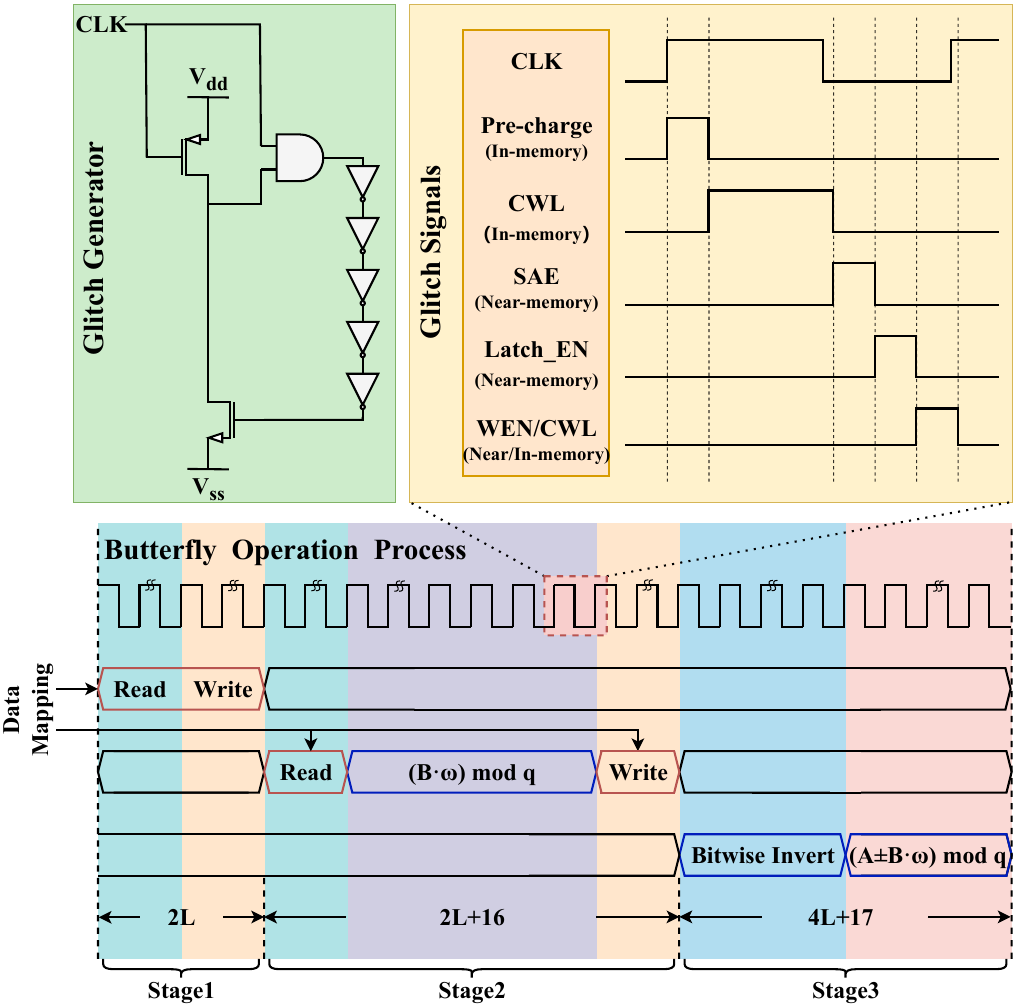} 
    \caption{Schematic of the glitch generator and critical timing waveforms for NTT operations.}
    \label{fig:glitch}
\end{figure}

The Glitch Generator, as shown in the top-left corner of Fig.~\ref{fig:glitch}, produces multiple narrow pulse-width signals within a single clock cycle, primarily used for controlling SRAM or near-memory computing units. Under the control of the glitch signals, near-memory computing units perform consecutive operations within a single clock cycle, significantly accelerating the speed of butterfly operations. Fig.~\ref{fig:glitch} illustrates the detailed schematic diagram of the glitch generator with its corresponding waveform, along with the timing diagram of a single-stage butterfly operation. The data mapping is completed through two data exchanges in $Stage 1$ and $Stage 2$. Note that a read or write operation requires $L$ cycles for $L$ bits of data. The modular multiplication takes 16 cycles. Bit reversal and addition/subtraction operations allow single-bit data to be read and processed within one cycle. During the implementation of addition/subtraction operations, the data designated for subtraction is first fully read out, subjected to bitwise inversion, and rewritten to the SRAM arrays. Subsequently, a parallel readout of all row data is executed to perform modular addition operations, thereby obtaining the final computational results. The addition/subtraction steps require almost no additional cycles except for the final modular operation. $Stage 3$ completes in 4$L$ + 17 cycles.

\section{Results and Discussion}
We evaluate the proposed NTT in a 28 nm CMOS process. Table \ref{comparison} provides the results and comparisons of recent NTT designs and this work. In the case of $N=256$, the throughput of the proposed NTT accelerator reaches 67.1 kNTT/s, with a minimal area overhead of only 0.006 $m^2$, achieving the highest throughput-per-area (11183.3 KNTT/s/mm²). In comparison, the proposed structure demonstrates 1.6×/2.9×/2.7x higher throughput than those of \cite{Zhang2023}, \cite{Nejatollahi2020} and \cite{ICASSP2020}, respectively, while achieving 4×/10× area reduction compared to \cite{Zhang2023} and \cite{Nejatollahi2020}. Although the NTT designs in \cite{Pakala2025}, \cite{Li2022}, \cite{Samardzic2021}, and \cite{song2018} outperform the proposed structure in throughput, they require at least 3.2× more area. This work demonstrates a balanced performance between throughput and area efficiency, providing 1.5$\sim$28× better throughput-per-area compared to the state-of-the-art.

\begin{table*}[htbp]
    \centering
    \setlength{\tabcolsep}{2pt}
    \caption{Comparison with State-of-the-Art NTT Accelerators}
    \label{comparison}
    \renewcommand{\arraystretch}{1.5}
    \begin{tabular}{l c c c c c c c c c c}
        \toprule
        Design & Type & \multicolumn{1}{c}{N} & \multicolumn{1}{c}{Freq} & \multicolumn{1}{c}{Bit} & \multicolumn{1}{c}{Tput.} & \multicolumn{1}{c}{Latency} & \multicolumn{1}{c}{Energy} & \multicolumn{1}{c}{Area} & \multicolumn{1}{c}{Tput./area} \\
        & & & \multicolumn{1}{c}{(MHz)} & \multicolumn{1}{c}{width} & \multicolumn{1}{c}{(kNTT/s)} & \multicolumn{1}{c}{($\mu$s)} & \multicolumn{1}{c}{(nJ/NTT)} & \multicolumn{1}{c}{(mm$^2$)} & \multicolumn{1}{c}{(kNTT/s/mm$^2$)} \\
        \midrule
        \multirow{3}{*}{This work (28nm)} & \multirow{3}{*}{ } & 256 & 176 & 14 & 67.1 & 14.9 & 622 & 0.006 & 11183.3  \\ 
         &  \makecell[c]{SRAM\\CIM} & 512 & 163 & 14 & 51.5 & 19.4 & 714 & 0.011 & 4681.8  \\
         &  & 1024 & 148 & 14 & 37.3 & 26.8 & 766 & 0.021 & 1776.2  \\ \hline
        TVLSI 2025 ($28\mathrm{nm}^a$) \cite{Pakala2025} & \makecell[c]{SRAM \\ CIM} & 1024 & 167 & 32 & 62 & 16.1 & 126 & 0.154 & 403  \\ \hline
        DAC 2023 ($28\mathrm{nm}^a$) \cite{Zhang2023} & \makecell[c]{SRAM \\ CIM} & 256 & 3800 & 16 & 26 & 38.5 & 13 & 0.024 & 1082 \\ \hline
        TVLSI 2022 ($28\mathrm{nm}^a$) \cite{Li2022} & \makecell[c]{SRAM \\ PIM} & 1024 & 151 & 14 & 80 & 12.5 & 93 & 0.067 & 1195  \\ \hline
        MICRO 2021 ($28\mathrm{nm}^a$) \cite{Samardzic2021} & Pipelined & 8K & 2000 & 128 & 36k & 0.028 & 1.24M & 68.139 & 524  \\ \hline
        DAC 2020 ($28\mathrm{nm}^a$) \cite{Nejatollahi2020} & \makecell[c]{ReRAM \\ PIM} & 256 & 909 & 16 & 23.4 & 42.7 & 1007 & 0.059\textsuperscript{b} & 397 \\ \hline
        ICASSP 2020 ($28\mathrm{nm}^a$) \cite{ICASSP2020} & \makecell[c]{FPGA} & 1024 & 182.98 & 16 & 14 & 71.3 & 6135 & - & -   \\ \hline
        CICC 2018 ($28\mathrm{nm}^a$) \cite{song2018} & \makecell[c]{ASIC} & 256 & 267 & 14 & 2.7k & 0.4 & 17 & 0.685\textsuperscript{b} & 3910  \\
        
        \bottomrule
    \end{tabular}
\smallskip
\\
    \footnotesize
    \begin{minipage}{\textwidth}
        \textbf{Notes}: 
        a. The process nodes are normalized to 28nm CMOS Process for an apples-to-apples comparison with GD-NTT.
        
        b.  The areas are estimated by referring to work \cite{Zhang2023}.
    \end{minipage}
\end{table*}
\section{Conclusion}
This brief proposes a near-memory NTT accelerator that enhances polynomial multiplication for post-quantum cryptography. Our design combines reconfigurable SRAM, glitch-driven execution, and near-memory computing to optimize throughput and efficiency. The architecture overcomes traditional memory bottlenecks and latency issues. Evaluation results show that GD-NTT can achieve a significant improvement in throughput-per-area(up to 28×) over the latest ASIC and in-memory designs, offering an efficient solution for edge devices and advancing cryptographic hardware research.
\bibliographystyle{IEEEtran}
\bibliography{ref}
\end{document}